# Computational study of radiative rate in silicon nanocrystals: Role of electronegative ligands and tensile strain

Katerina Dohnalova Newell[1a)], Prokop Hapala[1,2], Katerina Kusova[3], Ivan Infante[4,5]

[1] *University of Amsterdam, Institute of Physics, Science Park 904, 1098 XH Amsterdam, The Netherlands*
[2] *Department of Applied Physics, Aalto University, P.O. Box 11100, 00076 Aalto, Espoo, Finland*
[3] *Institute of Physics, Academy of Sciences of the Czech Republic, Cukrovarnicka 10, Praha 6, Czech Republic*
[4] *Department of Theoretical Chemistry, Faculty of Sciences, Vrije Universiteit Amsterdam, de Boelelaan 1083, 1081 HV, Amsterdam, The Netherlands*
[5]*Department of Nanochemistry, Italian Institute of Technology, via Morego 30, 16163, Genova.*
a) corresponding author k.newell@uva.nl

**Abstract**
It is widely accepted that the properties of most semiconductor nanocrystals can be tuned by their core size, shape and material. In covalent semiconductor nanocrystal materials, such as silicon, germanium or carbon, certain degree of tunability of the properties can be also achieved by the surface ligands. In particular, covalently bonded ligand species on the surface of such a nanocrystal (i) contribute to the density of states of the core via orbital delocalization; (ii) might introduce strain via ligand-to-ligand steric hindrance and (iii) will cause charge transfer from/to the core. In this work we study all these effects on silicon nanocrystals (SiNCs). We analyze geometrically optimized ~ 2 nm SiNCs with electronegative organic ligands using density functional theory (DFT) simulations. We show that the radiative rate is enhanced by electronegative alkyl and fluorocarbon with respect to what is expected from quantum confinement effect, while bandgap remains unchanged. Also, we show that tensile strain caused by the ligand steric hindrance is detrimental to the rate enhancement, contrary to the positive effects of the more homogeneous tensile strain induced in pressure cell.

**Introduction**
Silicon is a covalent group IV semiconductor with excellent electronic properties, but its indirect bandgap hampers its use for light sources in optoelectronic and photonic applications. Indirect bandgap also leads to weak band-edge absorption, causing thin film silicon solar cells to be very inefficient absorbers. Hence, pathway to achieve enhanced transition rate is highly sought for. Almost 3-order-of-magnitude rate enhancement is achieved in silicon nanocrystals (SiNCs) solely by the means of quantum confinement [1-6], which enables relatively good-efficiency light emitting devices [5-11]. It is, however, not enough to achieve competitive emissive properties with respect to direct bandgap materials with typical radiative rates of the order of $10^{8-9}$ s$^{-1}$.
In addition to the above-mentioned indirect-bandgap SiNCs, typically of larger cores and/or capped with oxide or long alkyl chains and with the fundamental radiative rates of the order of $10^4$-$10^5$ s$^{-1}$, other solution processed SiNCs appeared in the literature with faster fundamental emission rates of $10^7$-$10^9$ s$^{-1}$ (for an overview, see e.g. [6,12]). Interestingly, a transition from faster to slower radiative rates upon exchange of alkyl capping by an oxide, and vice e versa - transition from slow to fast rate upon exchange of a oxide by an alkyl - was experimentally confirmed [13-15]. The SiNCs with faster radiative rates exhibit, without any exception, a surface covered by various organic ligands, with most commonly a C-linked organic ligands [14-19], but also some N-linked species [20,21]. The proposed mechanisms behind these fast rates are numerous, ranging from Nitrogen related sites [17], strong coupling of the excited core states with either phonons or surface states [18], the enhancement of direct-bandgap transitions [16,22,23], or a combination of the above [24].

Clearly, from the experimental point of view, SiNCs seem to be a very complex system, where seemingly similar conditions might lead to different reported results and even the magnitude of radiative rates in H-capped SiNCs was experimentally reported to be either due to an indirect-bandgap [25] or a direct-bandgap [26]. Currently, probably the only two points with widespread agreement and acceptance by the scientific community is the role of quantum confinement [25,27] and the importance of surface states [6, 17, 25, 28, 29] in influencing the optical properties, but without any consensus on the origin of fast emission rate in SiNCs. The fact that numerous effects determine the optical properties complicates computational studies intending to reproduce and understand experiments. In the past, as a result of the lack of powerful computational techniques, theoretical studies were focusing only on rather small (< 1.8 nm) Si quantum dots [30-34], whose atomistic interpretation might differ when the size is increased (2-3 nm). Even though density-function-theory (DFT) calculations are usually preferred, empirical tight-binding models have been employed to facilitate computations on more complex and larger systems [22,34]. These have shown that H passivation is found to be the most reliable reference system to study the effect of quantum-confinement [32]. However, perhaps in agreement with experiments, the computed optical properties were very sensitive to the surface chemistry [30,33]. Later, more complex passivating organic groups and larger Si cores started to be addressed also within the DFT framework [31, 33, 35, 36], demonstrating benefits of alkyl passivation for light emission [33, 34]. However, even within DFT, some contradictions emerged in the calculated radiative rates when compared to experiments. For example, a 1.4-nm-large particle sparsely passivated with short alkenes (–$C_3H_6$ with 29% surface coverage at 0 K) was reported to reach the a radiative rate of $10^8$ $s^{-1}$ [33], which is well inside the direct-bandgap regime, whereas a later calculation of a slightly larger 1.9-nm-large nanoparticle with short alkyl ligands (–$C_2H_5$ with 72% surface coverage at 0 K) yielded much lower fundamental radiative rates of $6\times10^4$ $s^{-1}$ [36], typical of indirect-bandgap SiNCs.

In the past few years, we experimentally prepared SiNCs passivated by shorter alkyl ligands (methyl, butyl) with enhanced fundamental radiative rates in the direct-bandgap-semiconductor regime ($10^8$-$10^9$ $s^{-1}$) [14,15,23,37], demonstrated switch between fast and slow rates upon exchanges of organic and oxide cappings [13-15] and attributed the observed enhancement to the electronegativity of C (with respect to that of Si and H) or generally an electrostatic field with a similar electronegative effect [14,22], or to tensile strain [23], using both DFT [23] and tight-binding [14,22] calculations.

In this work, to pinpoint further the origin of the enhanced radiative rates, we compare the roles of strain and electronegative ligand in a systematic theoretical study by DFT. To relate to the experiment in the best possible way, we generated model systems of the SiNC core with reasonable core size (2.1 nm) and we computed temperature-averaged radiative rates. Rather than providing an exhaustive list of possible surface terminations, the surface ligands and ligand sites are carefully selected to reflect separately the effects of (i) electronegative groups and/or (ii) tensile strain on the nanostructure. To be able to analyze the effects of electronegativity [14,22] and tensile strain [23] separately, we cap the surface of SiNCs only sparsely (~50%) to remove possible steric hindrance. Afterwards, we test the effect of shorter ligand (CH$_3$) and that of tensile strain by steric hindrance by introducing a higher degree of surface coverage (100%). Whereas the presence of shorter ligands is confirmed to cause tensile strain and to enhance the fundamental radiative rates, the presence of a higher surface coverage leads to a decrease in the radiative rate. Therefore, we conclude that tensile strain per se is not enough to enhance radiative rate and one need to distinguish between the inhomogeneous tensile strain caused by ligand steric hindrance and the homogeneous hydrostatic-deformation related tensile strain, which has been shown to enhance radiative rate [23].

**Results**
First, we validate our approach by simulating geometrically optimized SiNCs of sizes between 1 nm and 3 nm fully capped with hydrogen (for details see Supplementary Materials, Fig. S1). We find excellent agreement with the literature for the radiative rates [3,22,38]. The bandgap energies are shifted by a constant value of 0.5 eV, which is a known issue of ground state DFT [39]. Nevertheless, this is not an issue here, since we are interested in a comparative study with respect to hydrogen capped SiNCs and radiative rates.

For the ligand study, we choose a ~2 nm SiNC model, which has approximately 50% of its Si atoms on the surface, but still a large enough crystalline core (Fig. S1, bottom). Building on our previous work, where Carbon-linked ligands were suggested as ideal for the radiative rate enhancement [14,22], we choose alkyl ligands with increasing electronegativity obtained by substituting H atoms with an increasing number of fluorine atoms in the alkyl backbone structure: -C$_4$H$_9$, -C$_3$H$_6$-CH$_2$F, -C$_3$H$_6$-CF$_3$, -C$_2$H$_4$-C$_2$F$_5$, -CH$_2$-C$_3$F$_7$ and -C$_4$F$_9$ (Fig. 1a). Increasing electronegativity is confirmed by the analysis of the Mulliken population in cp2k [40], showing increased net charge transfer to the ligand (Fig. 1b). To avoid at first the competitive effects of tensile strain [23], we use 50% surface coverage by ligands (i.e. about 25% of all Si atoms are bonded to a carbon atom). The remaining 50% sites are capped by hydrogen. For all the ligands, we use the same bonding sites to prevent significant differences in symmetry, which could just by themselves lead to changes in the density of states (DOS) and hence also the radiative rate. The strain is monitored by analysis of the average Si-Si bond length of all the Si atoms of the geometrically optimized SiNC (Fig. 1b). Average Si-Si bond length is constant for all the large ligands, and only a small bond shrinkage is observed for the most electronegative ligands, possibly caused by the lowered electronic density in the core, rather than a compressive strain.

Molecular orbitals in real-space (x-y plane view) for the highest occupied molecular orbital (HOMO) and the lowest unoccupied molecular orbital (LUMO) are plotted in Figs. 1c and S4. The LUMO appear to be in general more symmetric and homogeneously distributed over the NC core than HOMO, which generally shows elliptical symmetry. While an occasional protrusions of the electronic wavefunction into ligand is visible in all the cases with C-linked organic ligand, we do not observe surface trapped states (such as observed for the highly electronegative oxygen capping [41]).

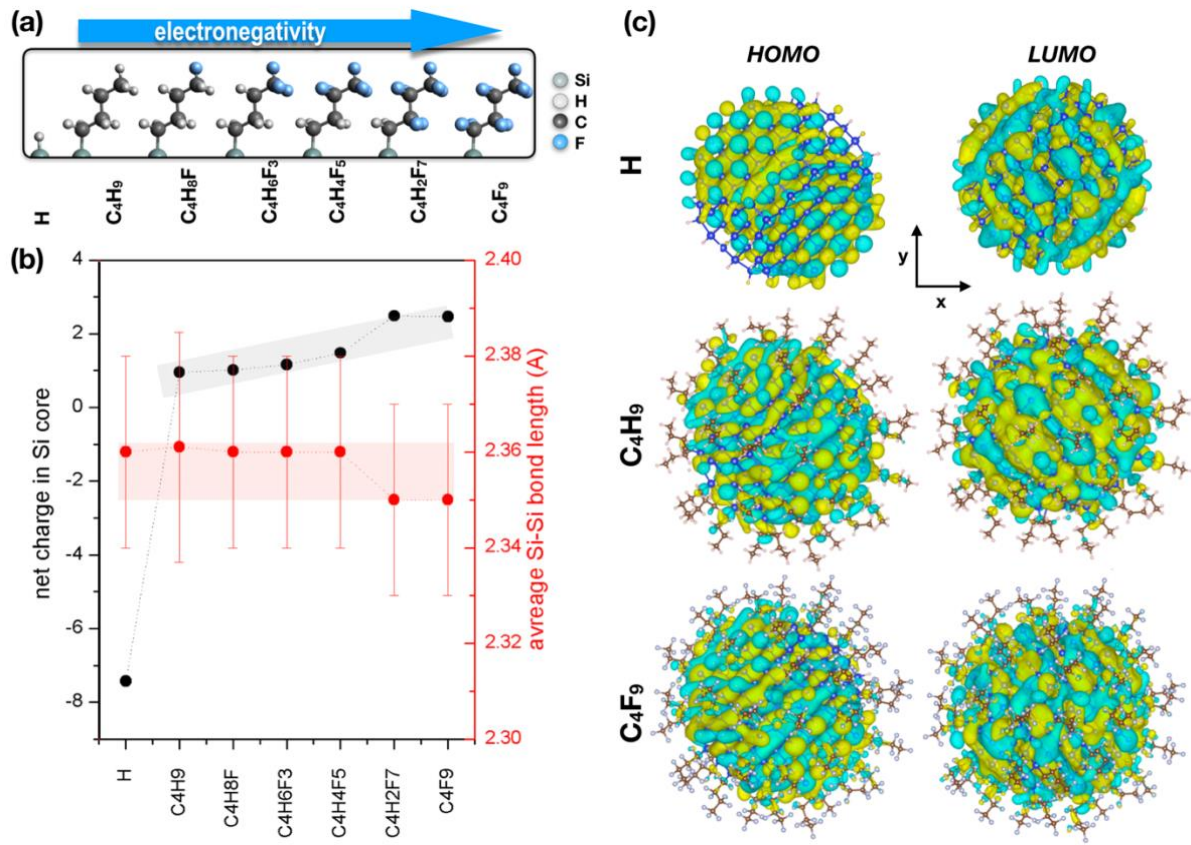

*Fig. 1* – *(a) Sketch of the SiNC ligands with increasing electronegativity. (b) Net charge and average Si-Si bond length in the SiNC (all Si atoms); colored lines are guide to the eye; Average bond data are obtained from histograms of Si-Si bonds achieved from the fully relaxed optimized SiNC structure, where the Si-Si bond lengths were fitted by a normal distribution (error bars are given by the standard deviation $\sigma$ here). (c) x-y plane view of the 3D real-space wavefunctions for HOMO and LUMO states for SiNC with the H, $C_4H_9$ and $C_4F_9$ capping (remaining systems are shown in Fig. S4).*

To understand the effect of electronegative ligands, we analyze the DOS as a function of energy and k-vector between the reciprocal-space Γ and X points (Fig. 2a). It is common to plot such a dependence in bulk crystals, however, in finite system like SiNC the k-vector is not a well-defined quantum number owing to a short-range translational symmetry. Nevertheless, using a Fourier transform of the real-space molecular orbitals and projecting them into the specific k-directions is possible and leads to a so called "fuzzy band-structure" (more details in [42]) that resembles for larger NCs the bulk band-structure and for very small NCs approaches that of an atom. This approach is especially valuable for indirect bandgap semiconductor NCs, such as SiNCs, where the k-space information is highly relevant. In brief, for H-capped SiNcs, such "fuzzy band-structures" were found to roughly follow the trends of the bulk material, while being more delocalized ("blurred") in reciprocal space due to the confinement of carriers (Fig. S2). As expected, long organic ligands lead to a lower localization of the carriers, resulting in a lower carrier confinement and therefore more bulk-like "fuzzy band-structure"(Fig. 2a) (i.e., with a smaller *k*-space "blur"). The next major observed feature observed for increasing electronegativity of the ligands is the lowered valence and conduction band edges, i.e. the decreasing Fermi energy (Figs. 2a,b), which results in the increasing work function for SiNC capped with more electronegative ligands. Despite these shifts, band-gap energy remains intact for most ligands (Fig. 2b). Nevertheless, apart from the significant differences between H and all the C-linked ligands, the DOS amongst the C-linked ligands appears to be very similar.

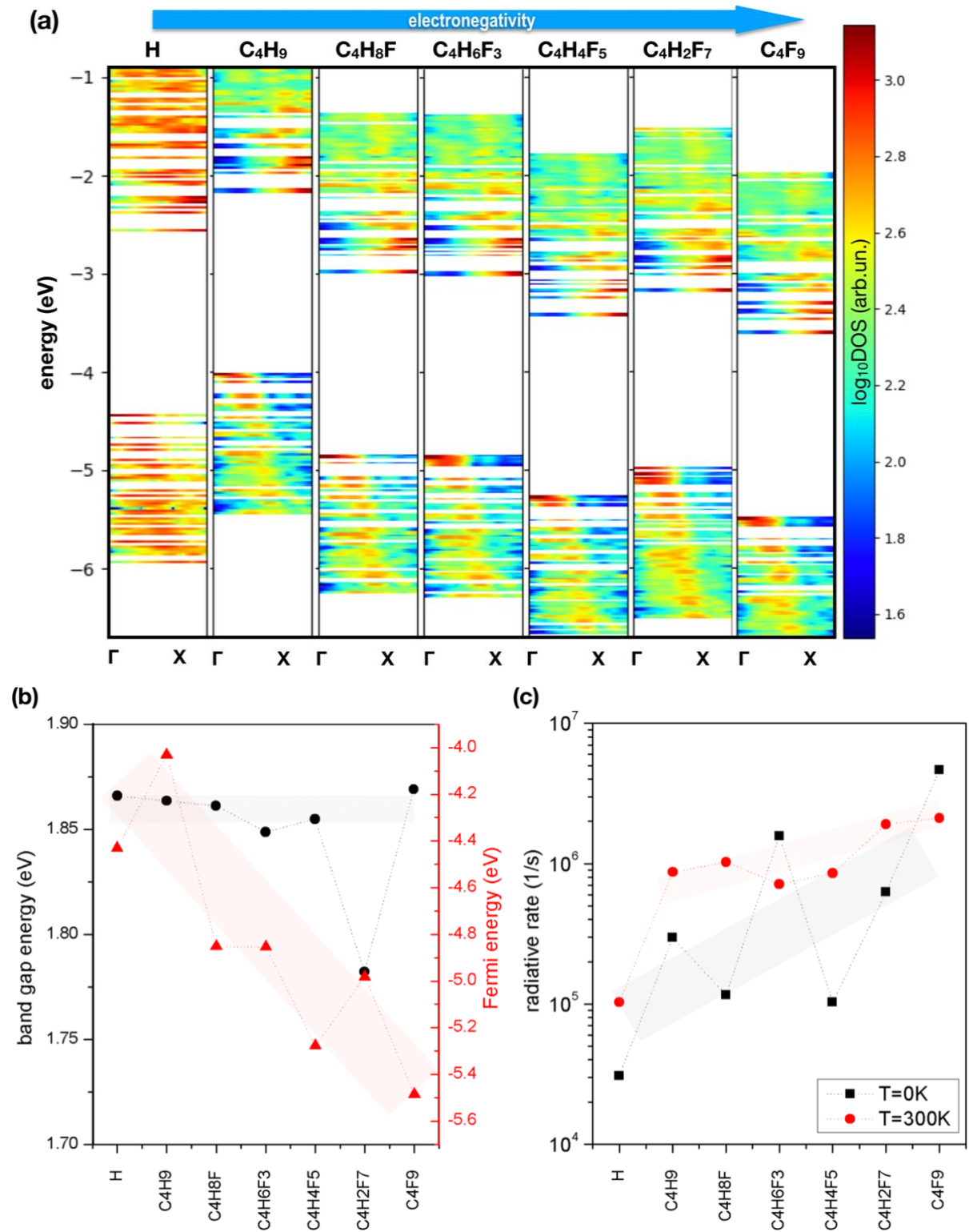

***Fig. 2*** *– 2.1 nm SiNC capped with ligands with increasing electronegativity: (a) "Fuzzy band structures"(logarithmic scaled DOS as a function of energy and k-vector in the Γ-X direction); (b) Band-gap energy (HOMO-LUMO transition) and Fermi energy; (c)Phonon-less radiative rate at 0K and thermalized at 300K. Colored lines in (b,c) serve as guide to the eye.*

To evaluate the effect of the electronegative ligand on the radiative rate, we evaluate the radiative rate $k_{rad}$ of the transition between an excited state $|j\rangle$ and the ground state $|i\rangle$ from the Einstein's coefficient of spontaneous emission:

$$A_{ij} = k_{rad} = \frac{1}{\tau_{rad}} = 2\frac{E_{ij}^2 f_{ij}}{c^3}, \qquad (1)$$

where $E_{ij} = E_j - E_i > 0$ is the energy of the transition, $c$ is speed of light and $f_{ij}$ is the oscillator strength given by Fermi's golden rule

$$f_{ij} = \frac{2m}{3\hbar^2} E_{ij} |d_{ij}|^2, \qquad (2)$$

where $m$ is mass of electron, $\hbar$ is the reduced Planck constant and $|d_{ij}|^2$ is the transition dipole moment, giving the strength of the interaction for the distribution of charge $q$ for the excited (electron) and ground (hole) states $d_{ij} = \langle j|qr|i\rangle$. Refractive index is assumed to be $n=1$ everywhere and is therefore omitted. For a more reasonable comparison with experiment, we evaluate also the thermally averaged rate, obtained from Eq. (1) and (2) by [43]

$$\langle k_{rad}\rangle = \frac{\sum_{i,j} k_{rad,ij} \exp(-\frac{E_{ij}}{k_B T})}{\sum_{i,j} \exp(-\frac{E_{ij}}{k_B T})}, \qquad (3)$$

where $k_B$ is the Boltzmann constant and $T$ is temperature.

The radiative rates calculated for the first 100 transitions are shown in Fig. S5. The lowest energy transition (the radiative rate at T=0K), given by Eq. (1), is extracted and plotted separately in Fig. 2c (black). The thermally averaged recombination rate for T=300K, given by Eq. (3) is plotted in Fig. 2c (red). We find that replacing hydrogen capping with 50% alkyl capping is beneficial for the radiative rate, leading to ~10 times enhanced rates. A slightly higher enhancement in the rate is achieved for the more electronegative ligands. This trend agrees with our previous findings [14,22], even though the enhancement is found to be smaller in this study.

Notable in Fig. S4 is also lowered degeneracy of the transitions: while the H and to a certain degree also the C$_4$H$_9$ and C$_4$H$_8$F capping keep the degenerate character of the lower energy transitions quite well, the more electronegative ligands lead to reduced degeneracy despite the fact that ligands are in all cases attached to the same surface sites.

To analyze the role of the "directness" of the HOMO-LUMO transition, we plot the DOS $k$-profile of the HOMO and LUMO states separately between the Γ and X points (from Fig. 2a), as shown in Figs. 3a,b. Contrary to our tight binding findings in [14,22], we do not find the direct correlation between the enhanced radiative rate and enhanced Γ contribution in the LUMO's DOS. In fact, only minor differences at the Γ point are observed in this study (see insets in Figs. 3a,b). However, interestingly, we see enhanced DOS contribution in the X HOMO's DOS, contributing to enlarged overlap of the e-h wavefunction in the X-X direct transition. Furthermore, when plotting the fraction of the DOS of the HOMO and LUMO states positioned on the ligand in real space, we find a slightly enhanced fraction of the LUMO and a bit decreased fraction of the HOMO on the ligands (Fig. 3c,d). This is expected also from the observed net charge transfer to the ligand (Fig.1b) caused by the higher electronegativity – where the hole (in the HOMO state) would retract to the SiNC core and the electron (in the LUMO state) would expand towards the ligand. A similar effect has been predicted also by our tight binding simulations in [14,22].

In summary, the effect of the electronegative ligands is a bit more complex than we suggested in [14,22], since one needs to consider the overall changes in the wavefunctions in both the k- and real-space. One more comment to the increased DOS in the LUMO Γ point is that a similar effect has already been used to qualitatively (i.e. without theoretical calculations) explain experimentally observed enhanced radiative rates in SiNCs with an N-linked ligand [24]. In that study, the authors argued that the increase of the Γ point LUMO DOS results from surface states. However, this is definitely not the case in the current study, because the Γ point LUMO DOS increase clearly does not correlate with higher wavefunction spatial localization on the surface. As can be seen from Figs. 3c,d and S7, the HOMO and LUMO states are localized in the cores of the SiNC for all the ligands.

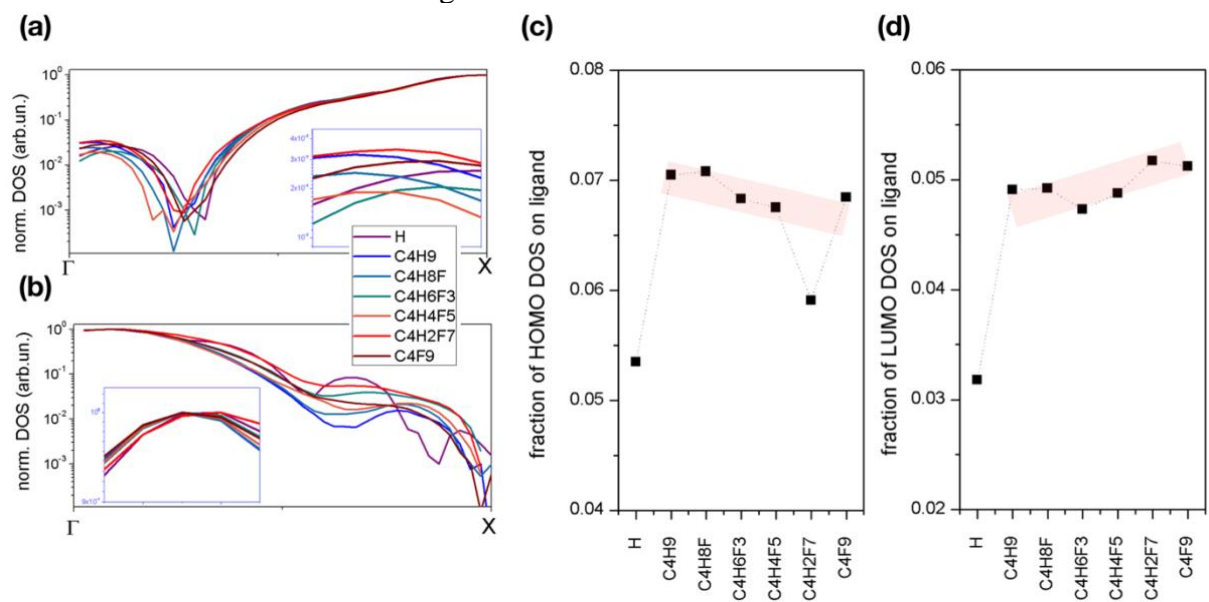

*Fig. 3* – *K-space profile of the DOS of the (a) LUMO and (b) HOMO DOS between the Γ and X points. Insets in (a,b) show detail in the Γ point. Fraction of the (c) HOMO and (d) LUMO DOS on the ligand; colored lines are guide to the eye.*

To test also the effects of strain induced by the steric hindrance proposed in Ref. 23, we analyze the *k*-space projected DOS (Fig. 4) and the radiative rates (Fig. S6) of the ~2nm SiNC (the same NC core as before) capped partially (50%) and fully (100%) by shorter capping by $-CH_3$ and $-CF_3$. We choose the short organic ligands in order to be able to relate to our previous work in Ref. 23. Ligands are placed to the same surface sites as previously used for longer ligands to prevent significant changes in symmetry and enable a direct comparison. First of all, we want to note that the short surface ligands do induce a small tensile strain in the nanocrystal, which is documented by the enlarged mean bond length (2.36 ± 0.02 Å in H-termined SiNC versus 2.37 ± 0.01 Å in a 50% $CH_3$ terminated SiNC, and 2.40 ± 0.04 Å in a 50 % $CF_3$ terminated SiNC; see Figure S7), which is not present in the SiNCs capped with longer butyl ligands with 2.36 ± 0.02 Å. Increasing the surface coverage of the shorter ligands from 50% to 100% then increases the mean bond length further (to 2.38 ± 0.02 Å for $CH_3$ terminated SiNC and 2.43 ± 0.05 Å in $CF_3$ terminated SiNC; see Figure S7). To pinpoint solely the effect of steric-hindrance-induced strain, we calculated the reciprocal-space-projected DOS and the radiative rates of the SiNCs capped with short ligands and present the results in Fig 4. One of the outcomes of the increased surface coverage is a red-shift in bandgap energy (although a relatively small one for the case of $–CH_3$ capping). Also, we see that the thermally averaged radiative rate (T= 300 K) for the partial covered SiNC with shorter ligands is lower than that of longer ligands ($5.9 \times 10^5$ s$^{-1}$ for $CH_3$ capped SiNC and $8.7 \times 10^5$ s$^{-1}$ in $C_4H_9$ capped SiNC as well as $3.1 \times 10^5$ s$^{-1}$ for $CF_3$ capped SiNC and $2.1 \times 10^6$ s$^{-1}$ in $C_4F_9$ capped SiNC). Furthermore, the rate is further decreased when increasing the capping percentage from 50% to 100% (from $5.9 \times 10^5$ s$^{-1}$ for partial $CH_3$ to $1.1 \times 10^5$ s$^{-1}$ for the full capping and from $3.1 \times 10^5$ s$^{-1}$ for partial $CF_3$ to $1.4 \times 10^5$ s$^{-1}$ for the full capping). The most straightforward interpretation of these results is that tensile strain induced by ligands is detrimental to the radiative rate. This is somewhat in contrast to our previous findings in Ref. 23, where we did not separate between the tensile strain caused by ligands (such as $CH_3$), which has detrimental effect, and by hydrostatic deformation, which appears to be generally favorable for the radiative rate enhancement [23]. To illustrate this, we can compare a 50% -$CF_3$ terminated and a 50% -$CH_3$ terminated SiNC, where the former clearly constitutes much more tensile strain because of its much larger mean bond length, at the same time leading to highly inhomogeneous straining, documented by the widened bond-length histogram (Fig. S7). This inhomogeneity of strain consequently causes the lowering of the radiative rates when compared to the less-strained $CH_3$ terminated SiNC as a result of a lowered real-space wavefunction overlap. The widening of the bond-length histogram consistently causes real-space localization on the LUMOs at the surface, in contrast to the partial capping and hydrogen termination, see Fig. S4.

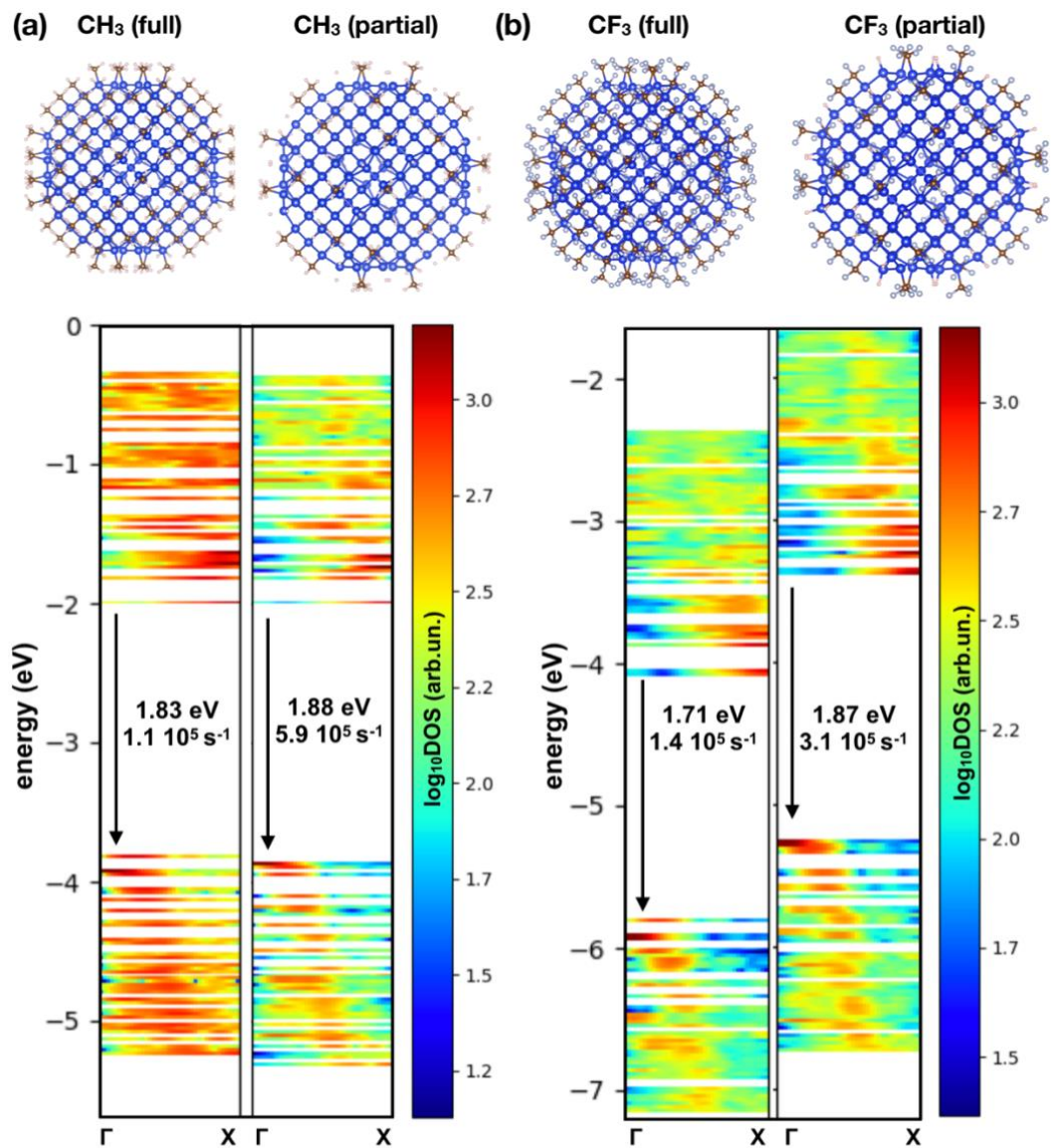

***Fig. 4*** – *Effect of steric hindrance in -$CH_3$ (a) and -$CF_3$ (b) capped SiNC on the DOS for partial (50%) and full (100%) surface coverages. Thermally averaged radiative rates and bandgap energies are indicated in the "fuzzy band-structures".*

Thus, tensile strain enhances radiative rates only when it acts homogeneously on the SiNCs. Interestingly, this observation is in accordance with the fact that the highest calculated radiative rates of alkyl-capped SiNCs were actually reported on "sparsely" capped SiNCs (up to 29% surface coverage) [33]. Moreover, it also well agrees with the theoretical calculations of our previous study [23], where additional 'external' hydrostatic strain had to be applied to the SiNC model in order to reach direct bandgap and explain the experimentally observed enhancement of radiative rates. Within the results of the present theoretical study, a plausible explanation is that even lower –$CH_3$ surface coverage would lead to a further increase in calculated radiative rates and, moreover.

In summary, we have shown that capping of SiNCs with carbon-linked electronegative ligand leads to an increase in the HOMO-LUMO radiative rate when compared to an H-terminated SiNC, even though not as high as previously suggested [14,22]. We specifically test the separate influences of (i) electronegativity and (ii) ligand-induced tensile strain on the fundamental radiative rate. We find that with increasing the electronegativity of the ligand, achieved by increasing the number of fluorine atoms on the ligand, (i) net charge is transferred towards the ligand; (ii) more of the LUMO's and less of the HOMO's DOS is located on the ligand; (iii) the Fermi energy decreases; (iv) the band-gap energy remains constant; and (v) the radiative rate of the HOMO-LUMO transition is enhanced. The presence of tensile strain is observed only in the shortest ligands and full surface coverage further increases it. The relationship between strain and enhanced radiative rates is not straightforward, though, because forcing a 100% surface coverage of the short methyl ligands causes a red-shift in bandgap-energy and is clearly detrimental to the radiative rates as a result inhomogeneity of the induced tensile strain, despite rising the overall tensile strain value. Thus we conclude that one needs to discern between inhomogeneous strain by ligands and homogeneous strain caused by hydrostatic deformation, which we have shown in past has a positive effects [23]. Because of the importance of the homogeneity of the induced strain, lower surface coverage seems to be favorable. Therefore, we suggest that sparse capping with short electronegative ligands might be the best way towards enhanced luminescence of SiNCs. The experimentally observed effect of enhanced radiative rates thus arises from a complex combination of several features, rather than only a direct-like band-gap formation, which is not explicitly observed here. The effect appears to be correlated with the net charge transfer from the core towards the ligands, without forming a surface trapping state for the case of butyl capping.

**Computational details**
Calculations were performed using DFT as implemented in the cp2k package Quickstep [40,44], with generalized gradient approximation (GGA) functional Perdew Burke Ernzerh (PBE) [45]. The core is approximated by the Goedecker-Teter-Hutter (GTH) pseudopotential potential [46]. For the basis set, we use short-range Gaussian double-zeta valence polarized basis set DVZP-MOLOPT [40]. Plane wave cutoff of integration grid was set to 400 Ry and self-consistency convergence to $10^{-6}$.

**Acknowledgments**
Authors acknowledge FOM Projectruimte No. 15PR3230 (KDN) and MacGillavry Fellowship from University of Amsterdam (KDN), Czech Science Foundation funding, Grant No. 18-05552S (KK) and support by the Operational Programme Research, Development and Education (Project No. SOLID21 CZ.02.1.01/0.0/0.0/16_019/0000760) (KK).

## Supplementary Materials

### Silicon nanocrystals (SiNCs) preparation and geometry

Si crystalline core is built in Vesta [1] editor by making a cut-off from a bulk crystal along the (100) (110) and (111) crystal planes. In this way, we obtain approximately spherical SiNCs of sizes 1 nm (Si35), 1.4 nm (Si87), 1.7 nm (Si147), 2.1 nm (Si263), 2.3 nm (Si377), 2.5 nm (Si405) and 3 nm (Si705) (Fig. S1). All SiNCs have crystalline core and the Si263 (2.1 nm in diameter) has approximately 50% Si atoms on the surface. The surface of the SiNCs has different types of facets with -$SiX_3$, -$SiX_2$ and -SiX bnding sites (Fig. S1). The -$SiX_3$ species are not chemically stable and are removed. The (100) facets are reconstructed (Fig. S1). Resulting Si nanocrystals are slightly tetrahedral and covered 100% by Hydrogen.

Alkyl and fluorocarbon ligands are attached in a free molecule editor Avogadro [2]. For partial surface coverage, we randomly select ~50% of the surface sites in such a way that Van der Waals spheres of the elements do not overlap. This is done in order to minimize unwanted effects of steric hindrance and from that arising strains. System is pre-relaxed in Avogadro and then fully relaxed using cp2k package Quickstep [3,4], until convergence is reached. Carbon-linked ligands are chosen with increasing electronegativity, indicated by increasing charge transfer to the ligand: -$C_4H_9$, -$C_3H_6$-$CH_2F$, -$C_3H_6$-$CF_3$, -$C_2H_4$-$C_2F_5$, -$CH_2$-$C_3F_7$ and -$C_4F_9$. Ligand surface coverage is 50% to avoid steric hindrance strain and the remaining sites are capped by Hydrogen.

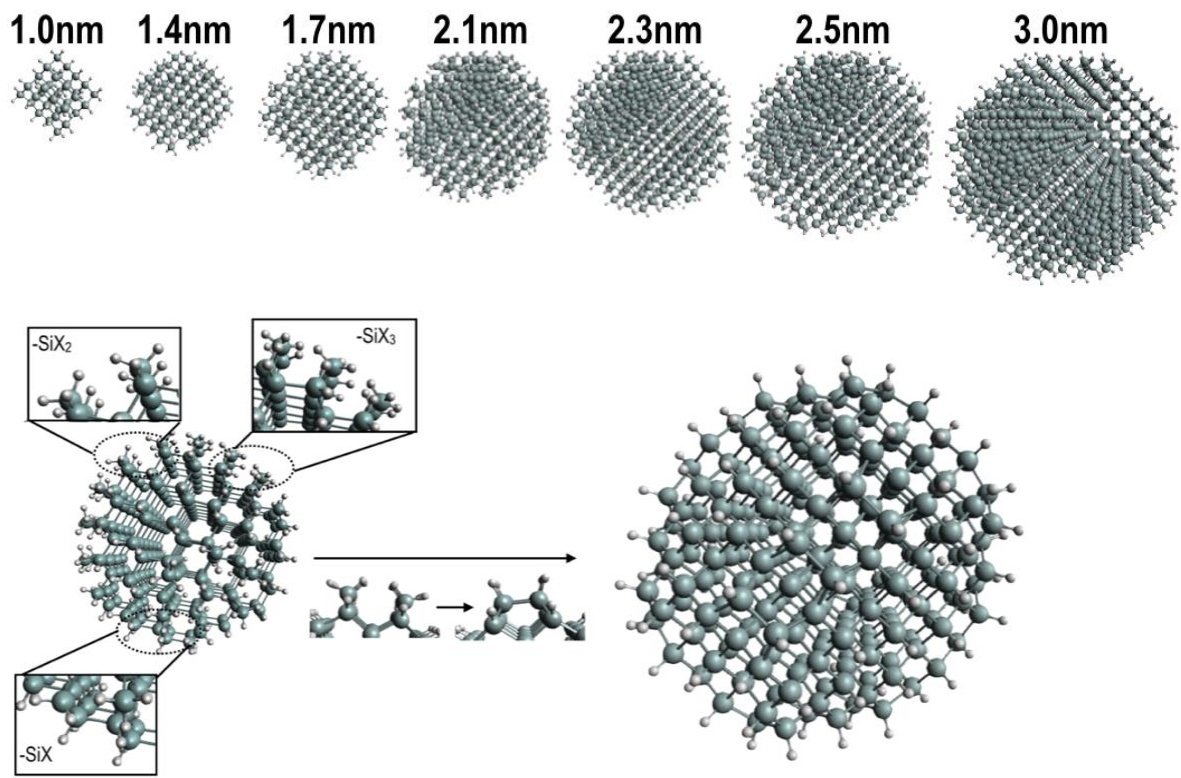

***Fig. S1*** *– (top) SiNCs of sizes between 1 and 3 nm, capped with Hydrogen. (bottom)From left to right: Surface sites on ~2nm nanocrystal Si263 and surface reconstruction of the edge (100) facets.*

**Size-dependent properties of H-capped SiNCs**

First we validate our approach by simulating hydrogen capped SiNCs of sizes between 1 and 3 nm (Fig. S1). Resulting 'fuzzy band-structures' (term coined in [5]), depicting density of states (DOS) as a function of energy and k-vector between Γ and X points in k-space, are shown in Fig. S2. Band-gap energies and radiative rates are separately plotted in Fig. S3.

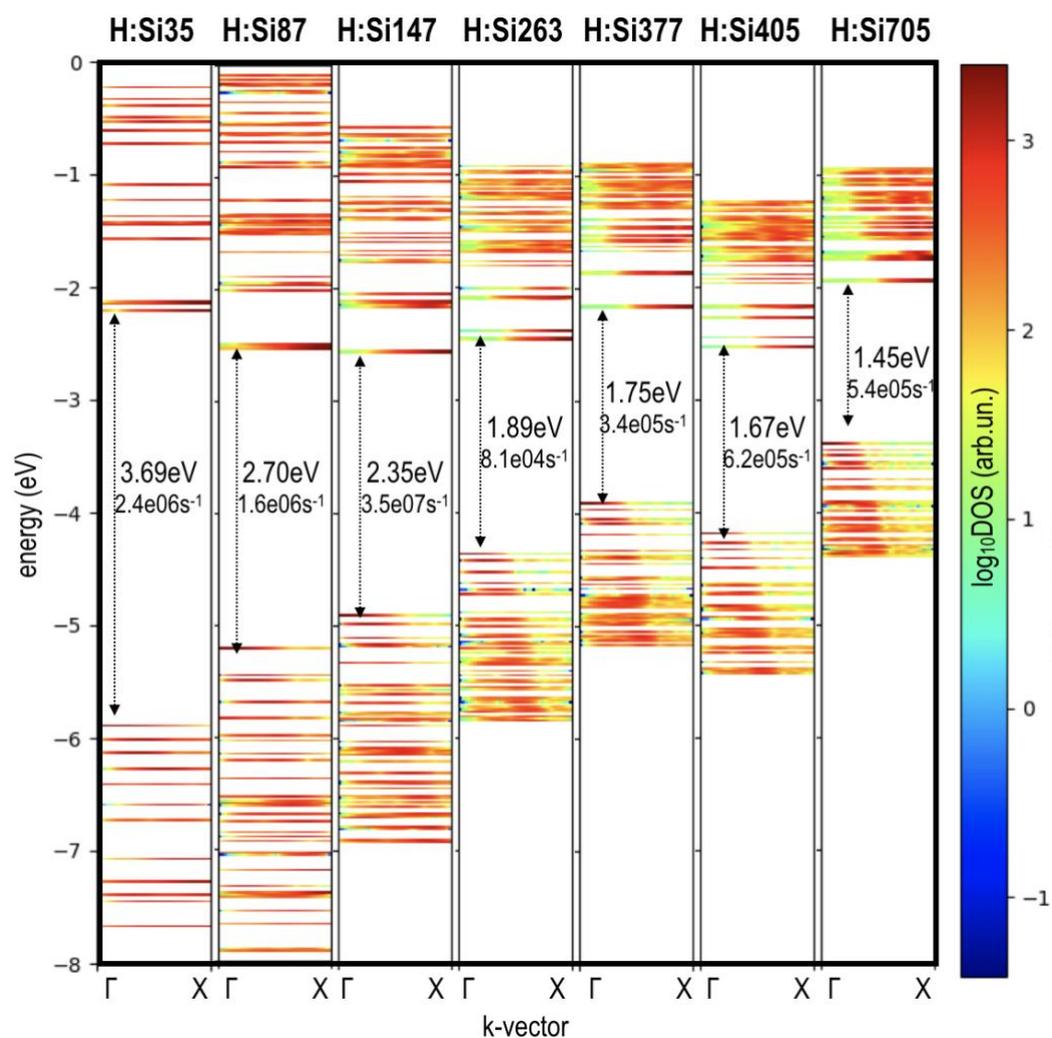

***Fig. S2*** – *DOS as a function of energy and k-vector in direction Γ - X of SiNCs between 1 and 3 nm, capped by Hydrogen. Bandgap energy and phonon-less radiative rate of the HOMO-LUMO transition is indicated inside each band structure.*

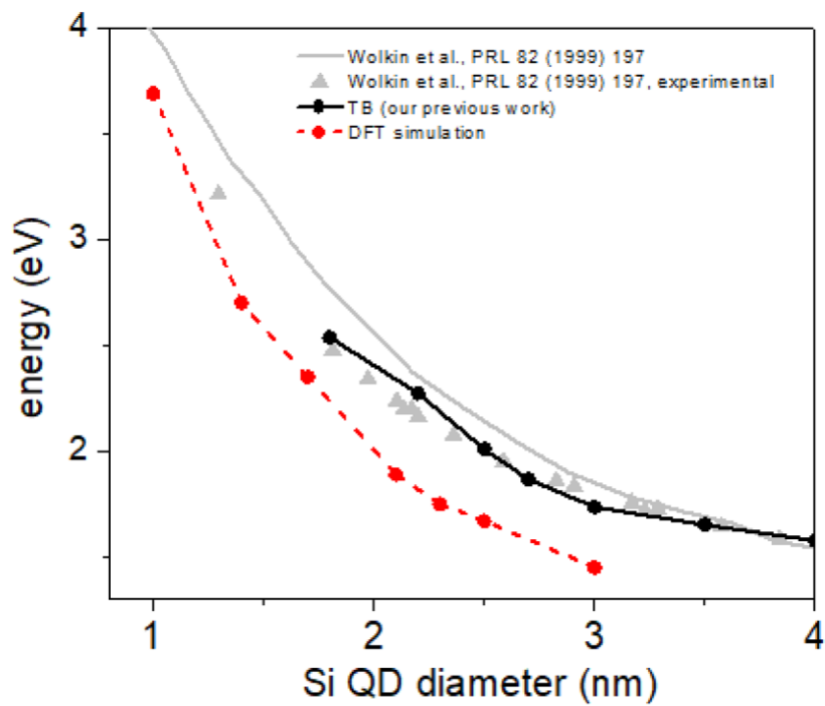

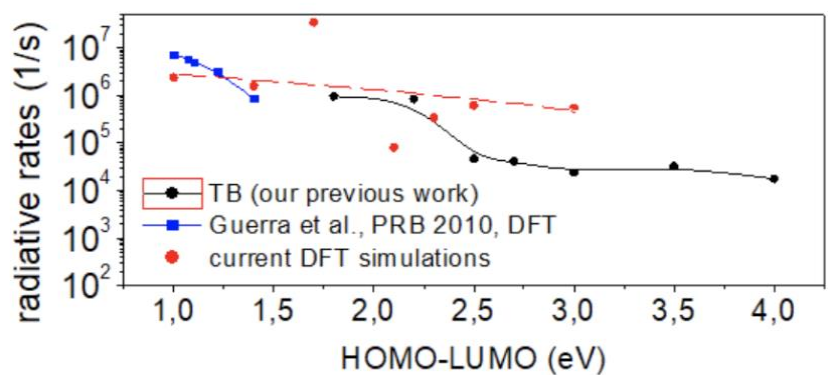

**Fig. S3** – *(top) Bandgap energy achieved by DFT in this work, compared to results in literature [6,7]. (bottom) Phonon-less radiative rate compared to literature [6,8].*

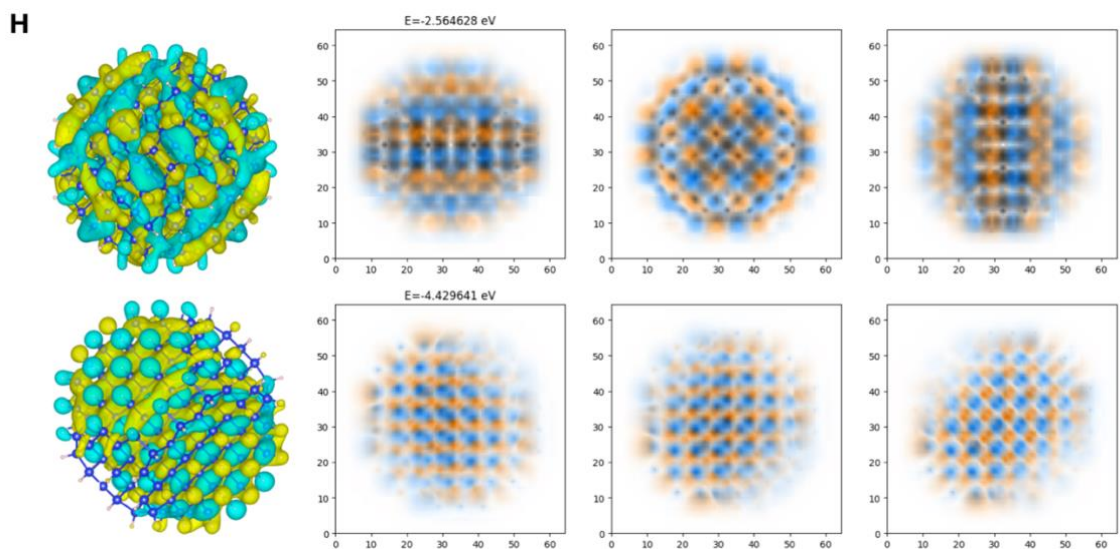

**C₄H₉**

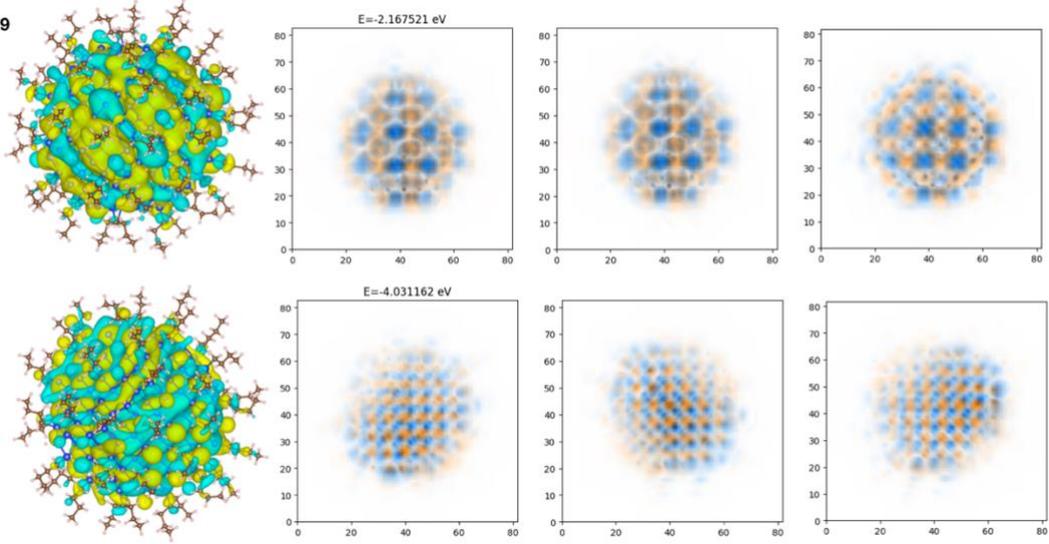

**C₄H₈F**

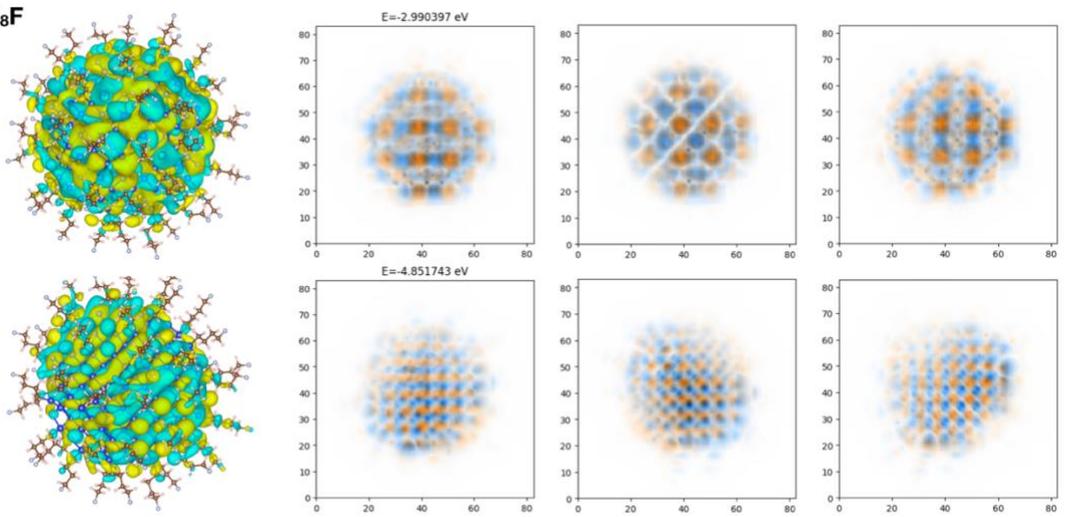

**C₄H₆F₃**

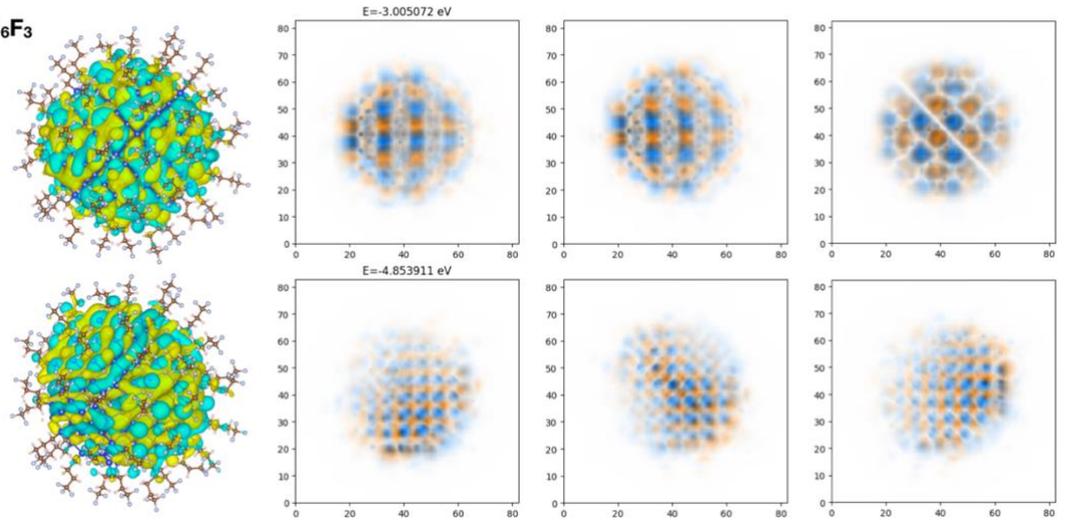

**C₄H₄F₅**

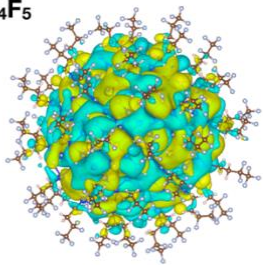
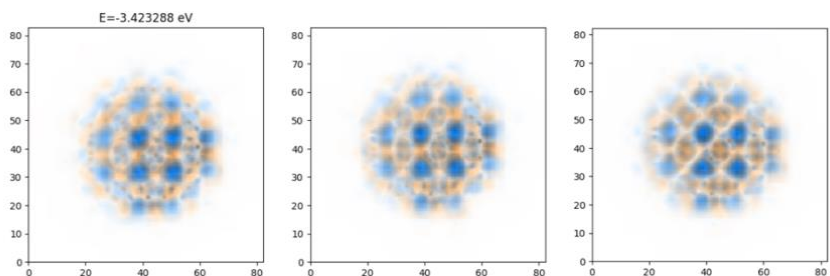

E=-3.423288 eV

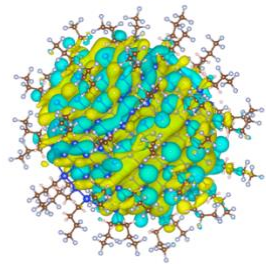
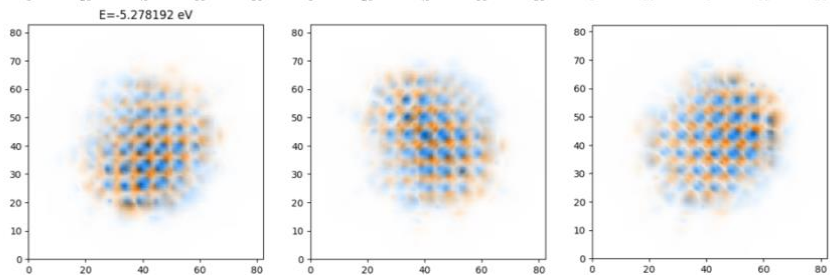

E=-5.278192 eV

**C₄H₂F₇**

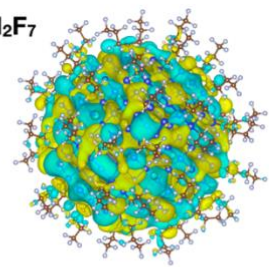
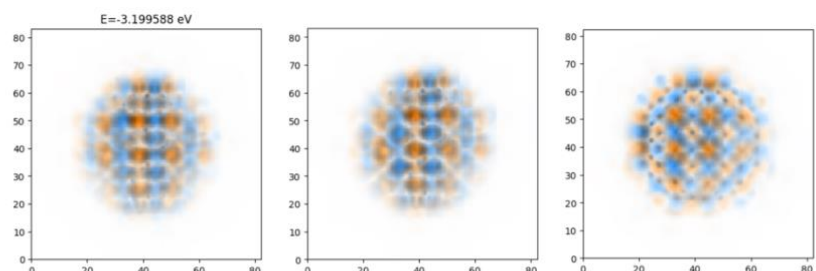

E=-3.199588 eV

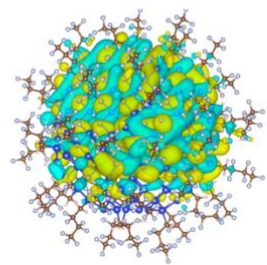
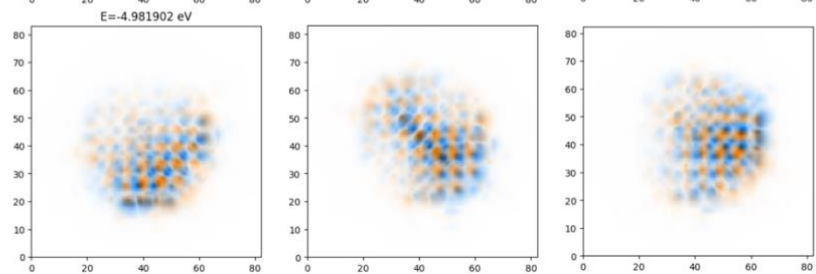

E=-4.981902 eV

**C₄F₉**

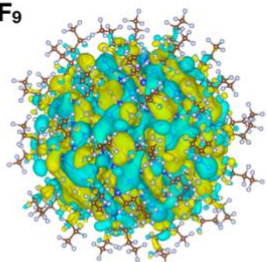
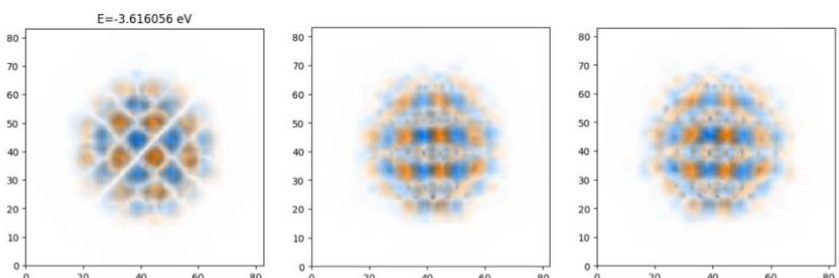

E=-3.616056 eV

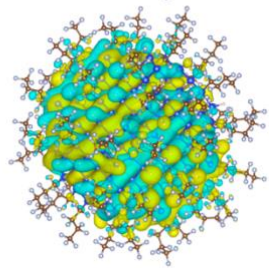
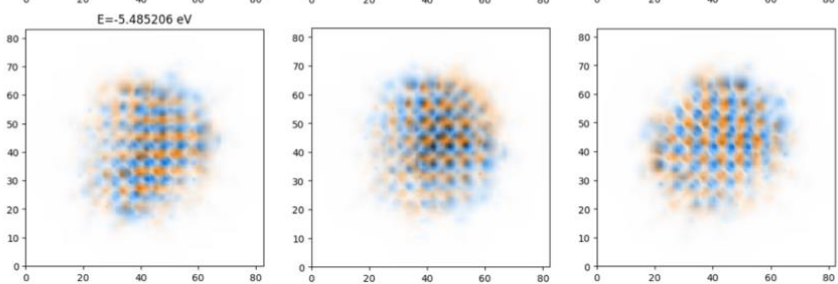

E=-5.485206 eV

**CH₃ (partial)**

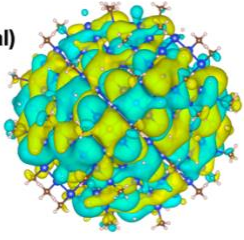
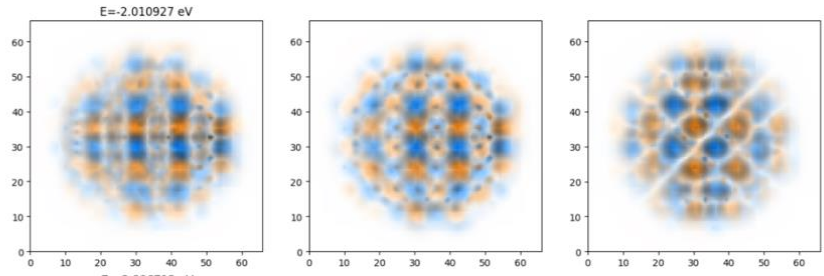

E=-2.010927 eV

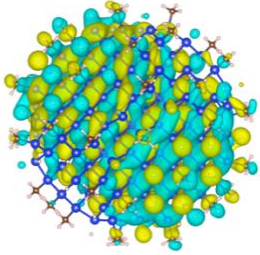
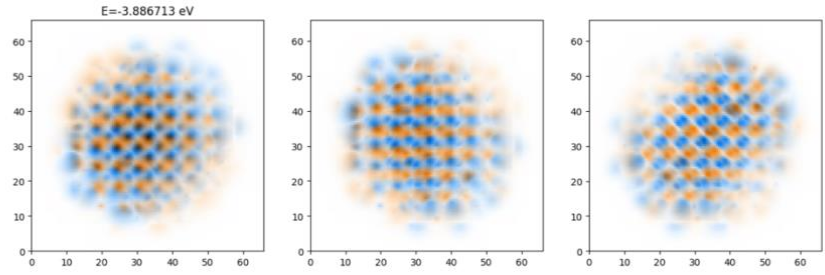

E=-3.886713 eV

**CH₃ (full)**

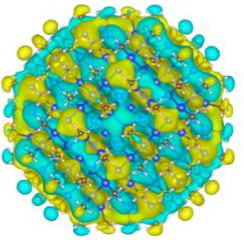
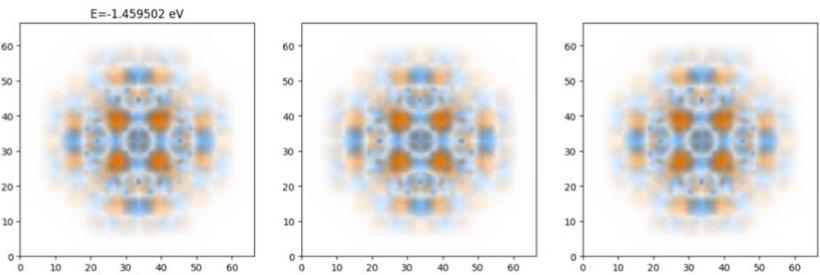

E=-1.459502 eV

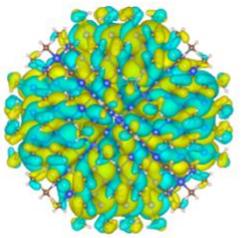
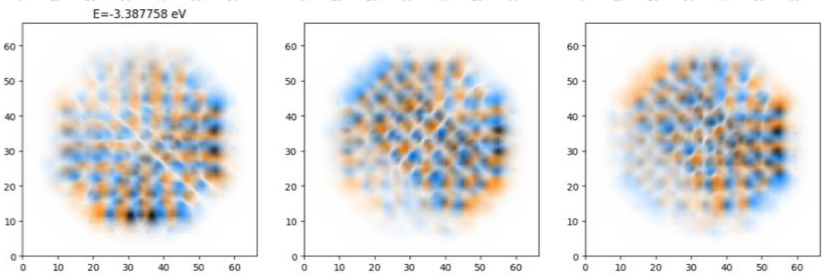

E=-3.387758 eV

**CF₃ (partial)**

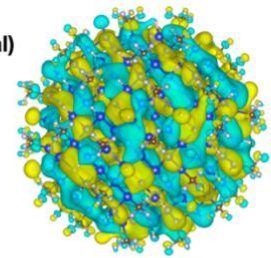
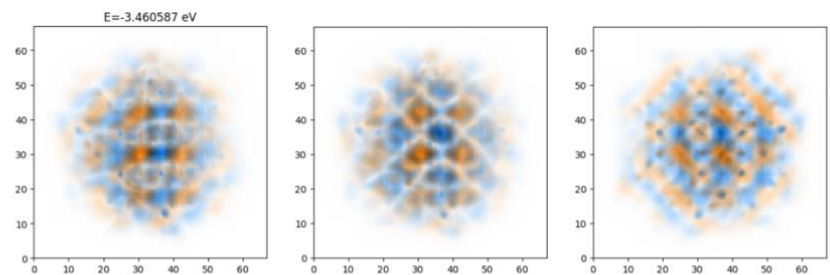

E=-3.460587 eV

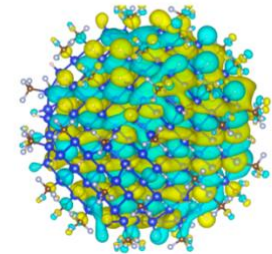
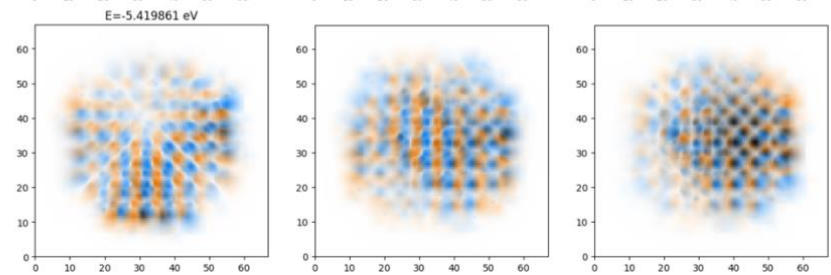

E=-5.419861 eV

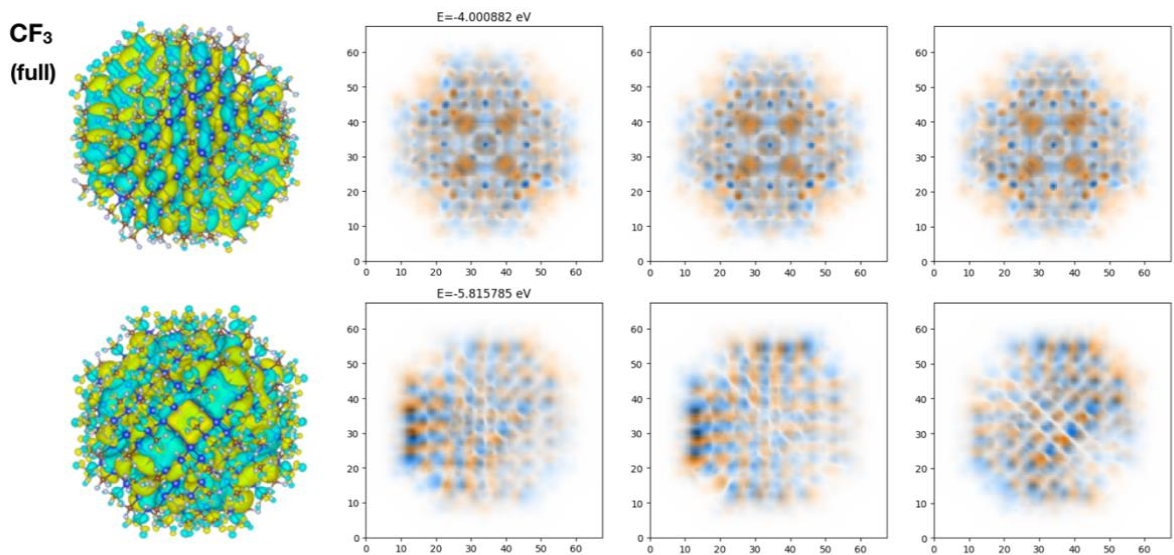

**Fig. S4** – *Real space HOMO (bottom rows, lower energy) and LUMO (top rows, higher energy) states for each SiNC-ligand system separately – 3D illustration is viewed from the x-y plane (left columns) and as a cross-section in x-z, x-y and y-z planes (3 right columns, from left to right).*

**Radiative rates in ligand-capped SiNCs**

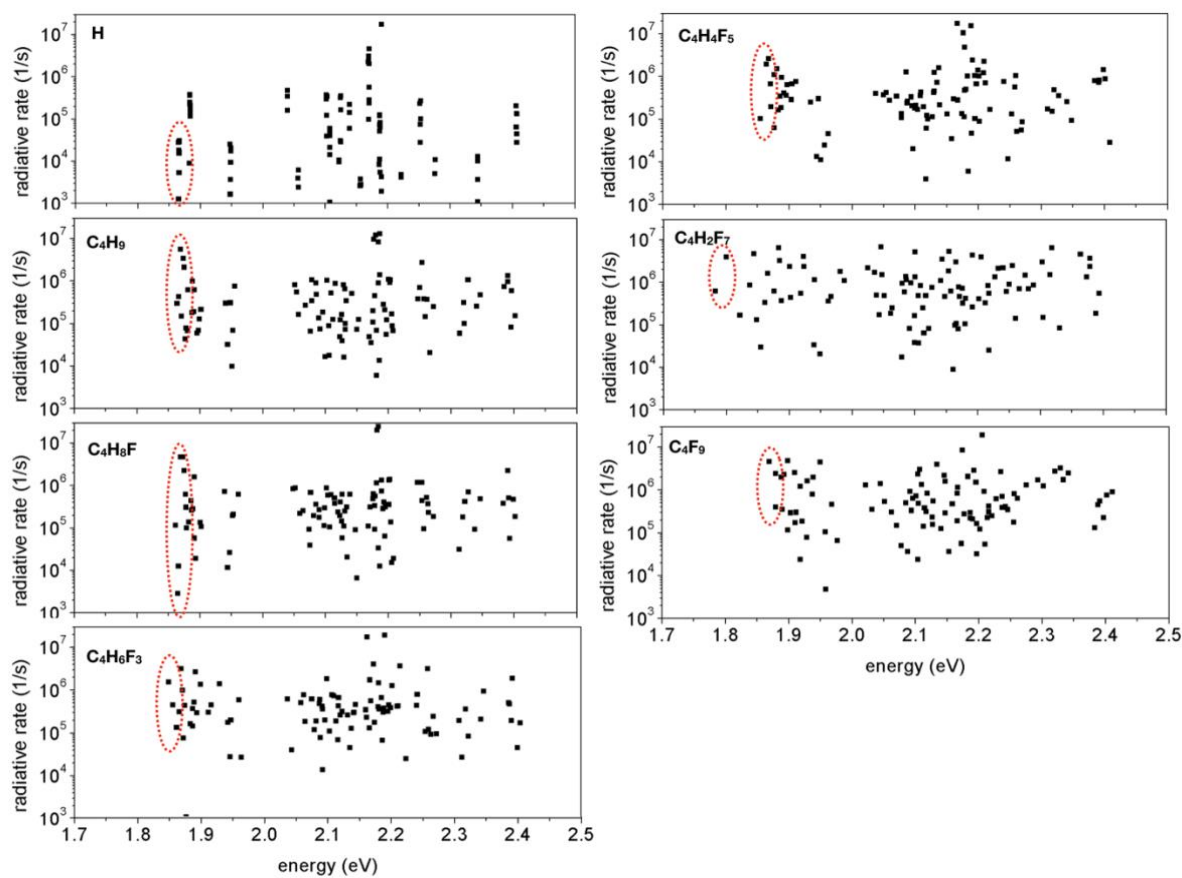

**Fig. S5** – *Radiative rates as a function of the transition energy. Lowest energy transitions are indicated by the red oval.*

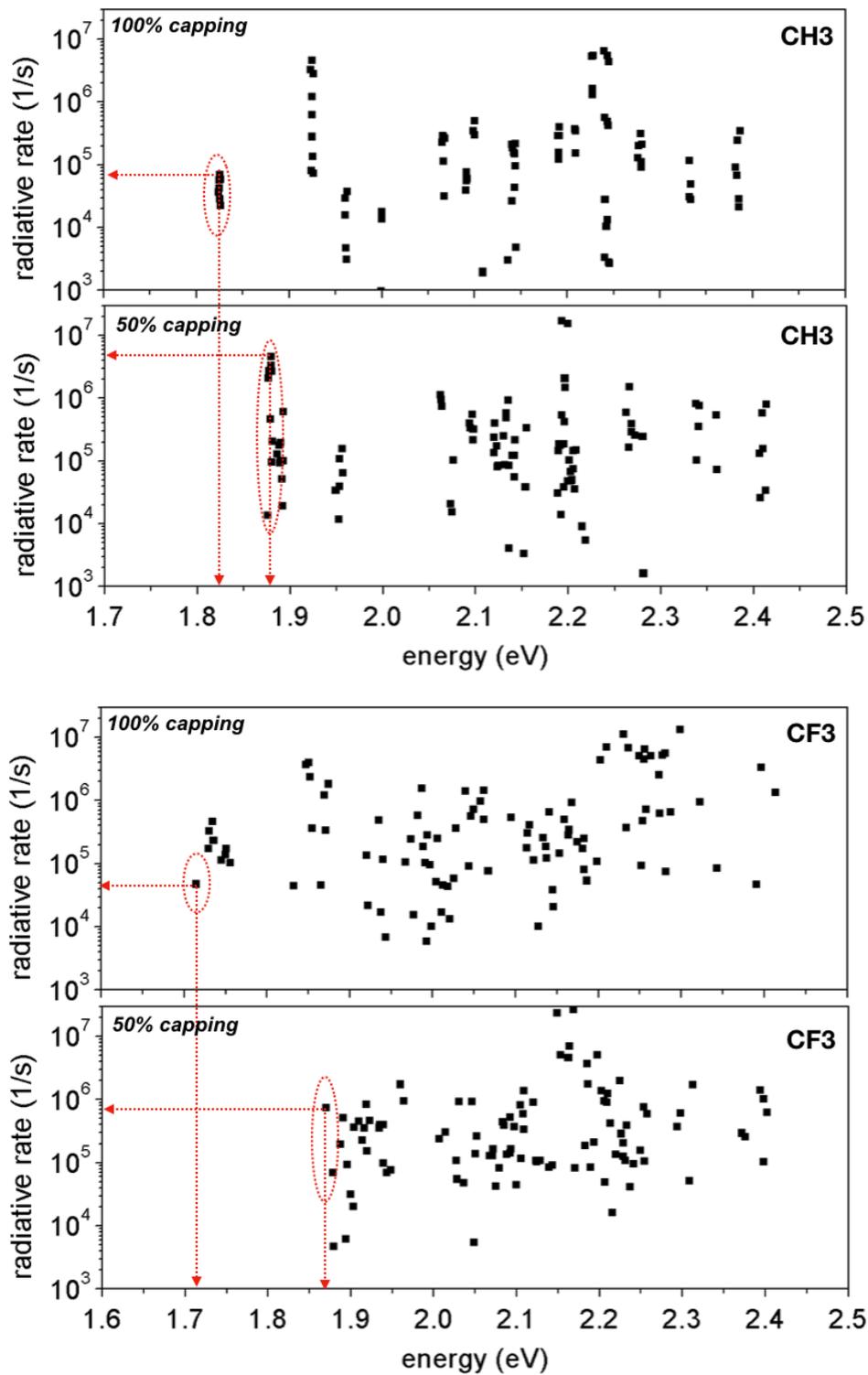

**Fig. S6** – *Radiative rates for the lowest 100 transitions in -CH$_3$ and -CF$_3$ capped 2nm SiNC (with 50% and 100% surface coverage). Lowest energy transitions are indicated by the red oval.*

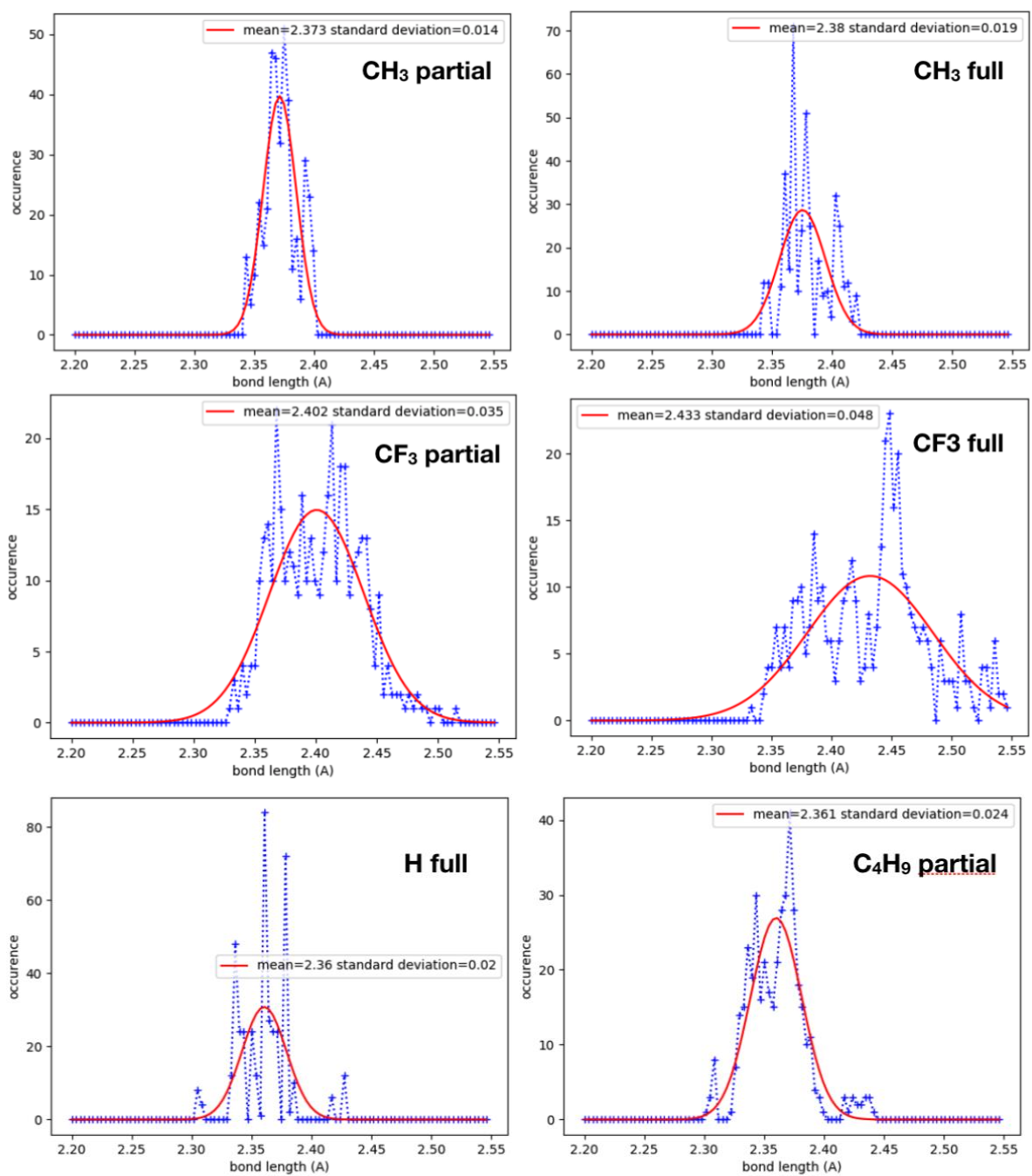

**Fig. S7** – *Si-Si bond length for SiNCs with partial (always 50%, rest is Hydrogen) and full surface coverage by H, CH₃, CF₃ and C₄H₉ ligands. Histograms are fitted by a normal distribution function, whose mean and standard deviations are shown in the figure legends.*